\documentclass[aps,twocolumn,epsfig,prb]{revtex4}

\newcommand{\bsigma}{\mbox{\boldmath $\sigma$}}
\newcommand{\bepsilon}{\mbox{\boldmath $\epsilon$}}
\newcommand{\bv}{\mbox{\boldmath $v$}}

\usepackage{amsmath}
\usepackage{graphicx}

\def\nn{\nonumber}

\begin{document}

\title{Polarization Dependence of Raman Spectra in Strained Graphene}

\author{Ken-ichi Sasaki}
\email[Email address: ]{SASAKI.Kenichi@nims.go.jp}
\affiliation{International Center for Materials Nanoarchitectonics, 
National Institute for Materials Science,
Namiki, Tsukuba 305-0044, Japan}

\author{Katsunori Wakabayashi}
\affiliation{International Center for Materials Nanoarchitectonics, 
National Institute for Materials Science,
Namiki, Tsukuba 305-0044, Japan}
\affiliation{PRESTO, Japan Science and Technology Agency,
Kawaguchi 332-0012, Japan}

\author{Toshiaki Enoki}
\affiliation{Department of Chemistry, Tokyo Institute of Technology,
Ookayama, Meguro-ku, Tokyo 152-8551, Japan}

\date{\today}
 
\begin{abstract}
 The polarization dependences 
 of the G, D, and 2D (G$'$) bands in Raman spectra
 at graphene bulk and edge are examined theoretically.
 The 2D and D bands have different selection rules 
 at bulk and edge.
 At bulk, the 2D band intensity is maximum when 
 the polarization of the scattered light is parallel to 
 that of incident light, whereas the D band intensity 
 does not have a polarization dependence.
 At edge, the 2D and D bands exhibit a selection rule 
 similar to that of the G band proposed in a previous paper.
 We suggest that a constraint equation 
 on the axial velocity caused by the graphene edge is
 essential for the dependence of the G band
 on the crystallographic orientation 
 observed in the bulk of strained graphene.
 This is indicative of that 
 the pseudospin and valleyspin in the bulk of graphene 
 can not be completely free from the effect of surrounding edge.
 The status of the experiments on the G and D bands
 at the graphene edge is mentioned.
\end{abstract}

\maketitle

\section{introduction}

Since the early stage of the research on graphene, 
characterization of a sample has been the central issue
and Raman spectroscopy has been playing a major role in 
characterizing a sample.~\cite{mildred10}
For example, the 2D (G$'$) band in Raman spectra is useful in 
distinguishing a monolayer 
from few-layer graphene stacked in the Bernal
configuration,~\cite{graf07,ferrari06,ni07}
and the appearance of the D band 
indicates that an intervalley elastic scattering of 
a photo-excited electron is activated by
defect.~\cite{thomsen00,saito03,malard09} 
Another advantage of Raman spectroscopy,
besides the characterization of a sample,
is that Raman spectra can include 
detailed information on 
the wave function of the electron.

Raman process concerns with photon, phonon 
and their mutual interaction through the electrons.
Because the electron-photon and electron-phonon interactions 
in graphene are relevant to pseudospin and valleyspin,~\cite{sasaki08ptps}
the Raman spectra are capable of retrieving information on 
the pseudospin and valleyspin.
An interesting point here is that 
the pseudospin and valleyspin are sensitive to 
the presence of graphene edge.~\cite{sasaki10-chiral}
As a result, we can have a selection rule
specific to the graphene edge.
For example, it is known that 
only the armchair edge enhances the D band intensity
and that the intensity depends on the angle 
between the armchair edge and 
the polarization of incident (scattered) laser light.~\cite{canifmmode04sec,you08,gupta09}

In a previous paper,
we proposed a selection rule for the G band.~\cite{sasaki09,sasaki10-jpsj}
This selection rule states, for example,
that the Raman intensity is enhanced 
when the polarization of 
Raman laser is parallel (perpendicular) 
to the armchair (zigzag) edge.
This prediction has been supported by 
recent experiments of Cong {\it et al.}~\cite{cong10}
and Begliarbekov {\it et al.}~\cite{begliarbekov10}
Their experiments illustrate that 
the G band intensity exhibits 
the anomalous polarization dependence at graphene edges
which is different from the polarization dependence at the interior
(bulk).
Their results could be naturally explained 
in terms of the special behavior 
of the pseudospin and valleyspin 
near the edges of graphene.

In this paper, we explore selection rules 
for the D and 2D (G$'$) bands at bulk and edge.
Since the 2D band is a prominent peak in Raman spectra,
the selection rule must be useful 
in extracting more information on the pseudospin and valleyspin
from the Raman spectra.
In addition, 
we examine the G band in strained graphene
as the application of the selection rule for the G band.
It is known that strain splits
the G band into two subbands called ${\rm G}^+$ and ${\rm G}^-$,
and that the Raman intensity of each subband has
a crystallographic orientation dependence.~\cite{huang09,mohiuddin09} 
Our result suggests that 
the crystallographic orientation dependence
observed in the bulk of strained graphene 
is relevant to the selection rule of the G band
for the graphene edge.
The pseudospin and valleyspin in the bulk of graphene 
seem to be not completely free from the effect of surrounding edge.

This paper is organized as follows.
In Sec.~\ref{sec:selection}
we derive the selection rule of the G band 
for the graphene edge 
in a unified manner. 
In Sec.~\ref{sec:strain} 
we apply the constraint which is essential 
for the selection rule of the G band
to explaining the crystallographic orientation dependence 
of the G band Raman intensity observed in strained graphene.
The selection rules of the D and 2D bands 
are proposed in Sec.~\ref{sec:Dand2D}.
Discussion and summary are given in Sec.~\ref{sec:ds}.

\section{Quick Overview of Selection Rule for G band}\label{sec:selection}

In this section 
we reproduce the selection rule of the G band 
for the graphene edge 
obtained in a previous paper,~\cite{sasaki10-jpsj}
by employing an approach 
based on two velocities associated with 
the gauge fields for photon and phonon.
This new approach can help us to recognize 
strange similarity between the zigzag and
armchair edges. 
This similarity is represented by the condition
Eq.~(\ref{eq:con1}) or Eq.~(\ref{eq:con2}).
As we will show in Sec.~\ref{sec:strain},
this condition 
is necessary to 
explain recent experiments showing that 
the G band exhibits a polarization dependence 
on the crystallographic orientation 
of strained graphene.~\cite{huang09,mohiuddin09}

\subsection{Two Velocities}

Let us begin with the Hamiltonian 
including photon field ${\bf A}$ 
and phonon fields ${\bf A}^{\rm q}$
and $\phi$,~\cite{sasaki08ptps}
\begin{align}
 H = 
 \begin{pmatrix}
  \bsigma \cdot (\hat{{\bf p}}-e{\bf A}+{\bf A}^{\rm q}) & 
  \phi \sigma_x \cr
  \phi^* \sigma_x & 
  \bsigma' \cdot (\hat{{\bf p}}-e{\bf A}-{\bf A}^{\rm q})
 \end{pmatrix},
 \label{eq:Hami}
\end{align}
where 
$\hat{{\bf p}}=-i\nabla$ is the momentum operator,
$\sigma_a$ ($a=0,x,y,z$) is the pseudospin,
$\bsigma \equiv (\sigma_x,\sigma_y)$, and
$\bsigma' \equiv(-\sigma_x,\sigma_y)$.
The phonon field $\phi$ (${\bf A}^{\rm q}$)
gives rise to an intervalley (intravalley) scattering.
In Eq.~(\ref{eq:Hami}), 
we have adopted units in which 
$\hbar = 1$ and $v_{\rm F}=1$, 
and omitted the position dependence in the variables
${\bf A}^{\rm q}$ and $\phi$
because we are interested in the $\Gamma$ and K points phonon modes.

From Eq.~(\ref{eq:Hami}), we define 
two velocity operators, ${\bv}$ and ${\bv}^{\rm q}$, 
as follows:
\begin{align}
\begin{split}
 & 
 {\bv} \equiv -\frac{1}{e}
 \frac{\partial H}{\partial {\bf A}}
 = 
 \begin{pmatrix}
  \bsigma & 0 \cr
  0 & \bsigma'
 \end{pmatrix},
 \\
  & 
 {\bv}^{\rm q} \equiv
 \frac{\partial H}{\partial {\bf A}^{\rm q}}
 = 
 \begin{pmatrix}
  \bsigma & 0 \cr
  0 & -\bsigma'
 \end{pmatrix}.
\end{split}
\end{align}
The operator ${\bv}$ is nothing but 
the usual velocity operator that 
couples to an electro-magnetic gauge field ${\bf A}$. 
Note that an electromagnetic current is given by 
multiplying $-ev_{\rm F}$ and ${\bv}$ together,
and the unperturbed Hamiltonian 
is written as $H_0 = \bv \cdot {\hat {\bf p}}$.
The velocity ${\bv}^{\rm q}$ is distinct from ${\bv}$ 
for the sign in front of $\bsigma'$. 
It may be appropriate to call ${\bv}^{\rm q}$ 
an axial velocity because ${\bv}^{\rm q}$ has
some analogy to the axial current in quantum electrodynamics.
In graphene, the axial velocity ${\bv}^{\rm q}$ 
couples with a lattice deformation.
Note that phonon is one example of a lattice deformation
and a general lattice deformation such as ripples and edges
can be represented 
by an axial gauge field 
${\bf A}^{\rm q}({\bf r})$.~\cite{kane97,sasaki05,katsnelson08}

By using $\tau_a$ $(a=0,x,y,z)$ to represent the valleyspin,
we write the component of the velocity ${\bv}$ as 
\begin{align}
 \begin{split}
  & v_x = \sigma_x \tau_z, \\
  & v_y = \sigma_y \tau_0,
 \end{split}
\end{align}
and that of ${\bv}^{\rm q}$ as 
\begin{align}
 \begin{split}
  & v_x^{\rm q} = \sigma_x \tau_0, \\
  & v_y^{\rm q} = \sigma_y \tau_z.
 \end{split}
\end{align}
Note that these two velocities ${\bv}$ and ${\bv}^{\rm q}$ 
are related with each other 
via the pseudospin $\sigma_z$ as 
\begin{align}
 v_i\sigma_z = -i \epsilon_{ij} v_j^{\rm q},
 \label{eq:jqj}
\end{align}
where $i,j \in \{x,y\}$ and $\epsilon_{ij}$ is 
antisymmetric tensor satisfying
$\epsilon_{xy}=1$, $\epsilon_{yx}=-1$, 
and $\epsilon_{xx}=\epsilon_{yy}=0$.
In the following subsections,
by examining the effect of the zigzag and armchair edges
on the two velocities
${\bv}$ and ${\bv}^{\rm q}$,
we derive the selection rule for the G band. 

\begin{figure}[htbp]
 \begin{center}
  \includegraphics[scale=0.4]{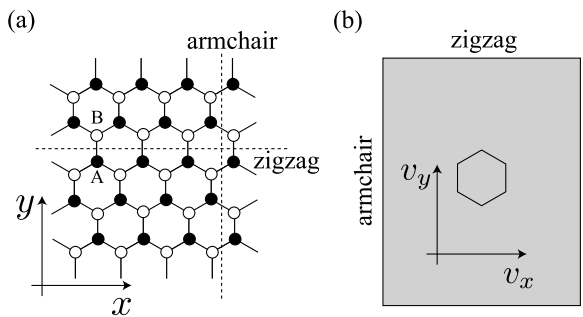}
 \end{center}
 \caption{(a) The coordinate system represents the crystal reference
 system. 
 The zigzag edge is parallel to the $x$-axis, while 
 the armchair edge is parallel to the $y$-axis.
 (b) A graphene system with rectangle shape enclosed by the zigzag and
 armchair edges. 
 }
 \label{fig:unit}
\end{figure}

\subsection{Zigzag Edge}

First of all, 
the electronic velocity ${\bv}$
normal to the zigzag edge must vanish.~\cite{mccann04,akhmerov08}
By taking the zigzag edge 
along the $x$-axis [see Fig.~\ref{fig:unit}(a)],
we thus have the condition 
\begin{align}
 \langle v_y \rangle = 0.
 \label{eq:cury}
\end{align}
Here, $\langle {\cal O} \rangle$ 
denotes the expectation value of the operator
${\cal O}$ with respect to 
the standing wave near the edge.
A general wave function can be represented by 
\begin{align}
 \Psi({\bf r}) = 
 \begin{pmatrix}
  \Psi_{\rm K}({\bf r}) \cr \Psi_{\rm K'}({\bf r})
 \end{pmatrix},
\end{align}
where $\Psi_{\rm K}({\bf r})$ [$\Psi_{\rm K'}({\bf r})$] is 
the two-component wave function 
for an electron near the K [K$'$] point.
The two-component structure corresponds to the pseudospin.
Note that the zigzag edge is not the 
source of intervalley scattering.~\cite{sasaki10-jpsj,sasaki10-forward}
As a result, the standing wave near the zigzag edge is written as
\begin{align}
 \Psi({\bf r}) = 
 \begin{pmatrix}
  \Psi_{\rm K}({\bf r}) \cr 0
 \end{pmatrix}, \ \ {\rm or} \ \ 
 \Psi({\bf r}) = 
 \begin{pmatrix}
  0 \cr \Psi_{\rm K'}({\bf r})
 \end{pmatrix}.
\end{align}
Thus, Eq.~(\ref{eq:cury}) leads to 
\begin{align}
\begin{split}
 & \int_S \Psi_{\rm K}^\dagger({\bf r}) \sigma_y 
 \Psi_{\rm K}({\bf r})dxdy=0, \\
 & \int_S \Psi_{\rm K'}^\dagger({\bf r}) \sigma_y 
 \Psi_{\rm K'}({\bf r})dxdy=0. 
\end{split}
 \label{eq:conzig}
\end{align}
Namely, the condition $\langle v_y \rangle = 0$ 
must be satisfied independently for the K and K$'$ points.
Because
\begin{align}
\begin{split}
 & (\tau_0 + \tau_z) \Psi({\bf r}) = 2 
 \begin{pmatrix}
  \Psi_{\rm K}({\bf r}) \cr 0
 \end{pmatrix}, \\
  & (\tau_0 - \tau_z) \Psi({\bf r}) = 2 
 \begin{pmatrix}
  0 \cr \Psi_{\rm K'}({\bf r})
 \end{pmatrix},
\end{split}
\end{align}
Eq.~(\ref{eq:conzig}) is possible only when 
$\langle \sigma_y (\tau_0 \pm \tau_z) \rangle = 0$ 
[i.e., $\langle v_y \rangle \pm \langle v_y^{\rm q} \rangle = 0$]
is also satisfied.
Hence, we get a constraint for the axial velocity as
\begin{align}
 \langle v_y^{\rm q} \rangle = 0.
 \label{eq:con1}
\end{align}
We will show below that 
the selection rule for the G band arises from 
the two conditions given by 
Eqs.~(\ref{eq:cury}) and (\ref{eq:con1}).

The Hamiltonian contains the electron-phonon (el-ph) interaction,
${\bf A}^{\rm q}\cdot {\bv}^{\rm q}$, and
the constraint Eq.~(\ref{eq:con1})
shows that $A_y^{\rm q}$ component 
does not have a nonzero el-ph matrix element.
Because the vector ${\bf A}^{\rm q}$
is pointing perpendicular to the 
optical phonon eigen vector,~\cite{dubay02,ishikawa06,sasaki10-jpsj}
$A_y^{\rm q}$ component 
corresponds to the optical phonon mode 
whose displacement vector is parallel to the
zigzag edge, $u_x$, which we called 
the longitudinal optical (LO) phonon mode
in a previous paper.~\cite{sasaki09,sasaki10-jpsj}
Thus, Eq.~(\ref{eq:con1}) shows that 
the LO mode is not a Raman active mode near the zigzag edge.~\footnote{
In the case of $\Gamma$ point phonon, 
the definitions of the LO and TO modes are not
unique since we do not have any reference vector. 
It seems standard that the LO mode is taken 
as the mode parallel with respect to the edge and the TO mode 
is the one perpendicular to the edge.}
Only transverse optical (TO) phonon mode 
$A_x^{\rm q}$ ($u_y$) can be Raman active.

Moreover, 
since the pseudospin $\sigma_z$ 
changes the wave function
from symmetric (bonding) 
to anti-symmetric (anti-bonding),
the optical transition amplitude 
is proportional to the expectation value of ${\bv}\sigma_z$.
The optical transition does not take place when the 
polarization of the incident (or scattered) laser light 
is parallel to the zigzag edge (or the $x$-axis) 
because the corresponding optical matrix element vanishes as
\begin{align}
 A_x \langle v_x \sigma_z \rangle = 
 -i A_x \langle \sigma_y \tau_z \rangle=
 -i A_x \langle v_y^{\rm q} \rangle = 0,
\end{align}
due to Eq.~(\ref{eq:con1}).
Similarly, the phonon softening (Kohn anomaly)
is relevant to the expectation value of ${\bv}^{\rm q}\sigma_z$.
The phonon softening is absent 
for the unique Raman active TO ($A_x^{\rm q}$) mode
because 
\begin{align}
 A_x^{\rm q} \langle v_x^{\rm q} \sigma_z \rangle 
 = -i A_x^{\rm q} \langle v_y \rangle = 0,
\end{align}
due to Eq.~(\ref{eq:cury}).
Note that the LO mode ($A_y^{\rm q}$) can undergo a phonon softening
effect, however, the LO mode is invisible to Raman spectra.

In summary, the polarization of the laser light should be 
perpendicular to the zigzag edge in order to have a Raman intensity,
and the corresponding Raman active mode is the TO mode which is free
from the phonon softening effect.
It is important to recognize that 
this selection rule is a consequence of the conditions
Eqs.~(\ref{eq:cury}) and (\ref{eq:con1}) for the electronic and axial
velocities.

\subsection{Armchair Edge}

The armchair edge can be examined in a manner similar 
to that for the zigzag edge. 
The electronic velocity normal to the armchair edge must vanish.
By taking the armchair edge 
along the $y$-axis (see Fig.~\ref{fig:unit}(a)), 
we have the condition 
\begin{align}
 \langle v_x \rangle = 0.
 \label{eq:cur2}
\end{align}
Note that the armchair edge 
is not the source of intravalley scattering
and preserves the pseudospin under 
an intervalley scattering.~\cite{sasaki10-jpsj,sasaki10-forward}
As a result, we obtain ($i=x,y$)
\begin{align}
 \int_S \Psi_{\rm K}^\dagger({\bf r}) \sigma_i 
 \Psi_{\rm K}({\bf r})dxdy=
 \int_S \Psi_{\rm K'}^\dagger({\bf r}) \sigma_i
 \Psi_{\rm K'}({\bf r})dxdy.
\end{align}
This equation is equivalent to 
the conditions $\langle \sigma_x \tau_z \rangle=0$
and $\langle \sigma_y \tau_z \rangle=0$.
The former condition $\langle \sigma_x \tau_z \rangle=0$
is nothing but Eq.~(\ref{eq:cur2}),
while the latter one $\langle \sigma_y \tau_z \rangle=0$
corresponds to 
\begin{align}
 \langle v_y^{\rm q} \rangle = 0.
 \label{eq:con2}
\end{align}
From this condition it is straightforward to show that
only the LO mode $A^{\rm q}_x$ ($u_y$) 
is Raman active mode at the armchair edge.
The optical transition does not take place when the 
polarization of the incident laser light is perpendicular 
to the armchair edge because 
\begin{align}
 A_x \langle v_x \sigma_z \rangle = 
 -i A_x \langle v_y^{\rm q} \rangle = 0,
\end{align}
due to Eq.~(\ref{eq:con2}).
Furthermore, we see by using Eq.~(\ref{eq:cur2}) that 
the Raman inactive TO mode ($A_y^{\rm q}$)
does not undergo the phonon softening
because 
\begin{align}
 A_y^{\rm q} \langle v_y^{\rm q}\sigma_z \rangle 
 = i A_y^{\rm q} \langle v_x \rangle = 0.
\end{align}
The Raman active LO mode ($A_x^{\rm q}$)
can exhibit a phonon softening effect.
To summarize the selection rule of the G band 
for the armchair edge, 
the polarization of the laser light should be 
parallel to the armchair edge in order to have a Raman intensity,
and the corresponding Raman active mode is the LO mode 
which undergoes a phonon softening effect.

The zigzag and armchair
edges are distinct concerning the usual velocity ${\bv}$ as 
$\langle v_y \rangle=0$ and $\langle v_x \rangle=0$, respectively.
Note, however, that 
the zigzag and armchair
edges are not distinct 
with respect to the axial velocity ${\bv}^{\rm q}$.
The constraint for the axial velocity at the armchair edge
is the same as that at the zigzag edge as shown by
Eqs.~(\ref{eq:con1}) and (\ref{eq:con2}), although
the origins of Eqs.~(\ref{eq:con1}) and (\ref{eq:con2})
are totally different.
For the zigzag edge, Eq.~(\ref{eq:con1}) is satisfied 
both for the K and K$'$ points since the zigzag edge
is not the source of an intervalley scattering.
In other words, the zigzag edge affects only the pseudospin
as $\langle \sigma_y \rangle=0$.~\cite{sasaki10-jpsj}
The zigzag edge is irrelevant to the valleyspin.
In contrast, 
Eq.~(\ref{eq:con2}) is satisfied because 
the armchair edge is the source of an intervalley scattering 
and preserves the pseudospin.
The armchair edge affects only the valleyspin and 
is irrelevant to the pseudospin.

\section{Uniaxial Strain}\label{sec:strain}

In this section 
we apply the selection rule of the G band for graphene edge 
to the G band in strained graphene.
It is known that strain lifts the degeneracy of the G band, 
so that the G band splits into two subbands called 
${\rm G}^+$ and ${\rm G}^-$.
The eigenvectors for the atomic displacements are perpendicular 
to the direction of strain for the ${\rm G}^+$ mode,
and parallel to the ${\rm G}^-$ mode.~\cite{huang09,mohiuddin09} 
By defining ${\bf A}^{\rm q}_{+}$ (${\bf A}^{\rm q}_{-}$)
field for the ${\rm G}^+$ (${\rm G}^-$) mode, 
we see that 
${\bf A}^{\rm q}_{+}$ is parallel to the direction of strain,
while ${\bf A}^{\rm q}_{-}$ is perpendicular to it
since ${\bf A}^{\rm q}$ is pointing perpendicular 
to the corresponding optical phonon eigenvector.
Recent experiments by Huang {\it et al.}~\cite{huang09} 
and Mohiuddin {\it et al.}~\cite{mohiuddin09}
for bulk of strained graphene show that
the Raman intensity of each ${\rm G}^+$ and ${\rm G}^-$ mode
has the crystallographic orientation dependence.
We suggest that 
the crystallographic orientation dependence 
of the G band Raman intensity 
observed at bulk 
can be attributed to the constraint 
$\langle v_y^{\rm q}\rangle =0$ 
obtained in the previous section for the graphene edge.
The difference between the bulk and edge 
is pointed out with respect to the crystallographic orientation
dependence of the G band Raman intensity
and Kohn anomaly effect.

\begin{figure}[htbp]
 \begin{center}
  \includegraphics[scale=0.4]{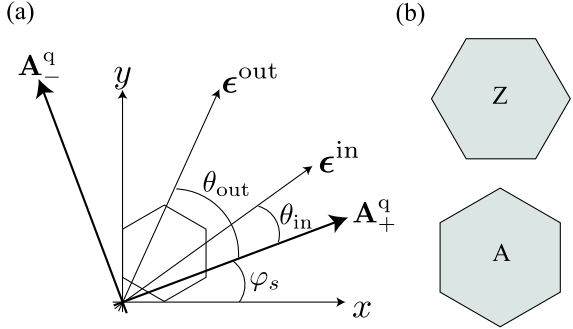}
 \end{center}
 \caption{(a)
 The strain is at the angle $\varphi_s$ with respect to the $x$-axis. 
 The direction of strain is parallel to ${\bf A}^{\rm q}_+$.
 The angle between the strain and the incident (scattered)
 laser polarization is denoted by $\theta_{\rm in}$ ($\theta_{\rm out}$).
 The axial vector ${\bf A}^{\rm q}_+$ corresponds to the ${\rm G}^+$
 mode, while ${\bf A}^{\rm q}_-$ corresponds to the ${\rm G}^-$ mode.
 (b) A hexagonal graphene with zigzag edges (top), and that with
 armchair edges (bottom).
 }
 \label{fig:strain}
\end{figure}

\subsection{Bulk}

Since the polarization of light, $\bepsilon$,
is parallel to ${\bf A}$, the electron-photon interaction,
${\bf A} \cdot \bv$,
is proportional to $\bepsilon\cdot \bv$.
Let the polarizations of incident light 
and scattered light be 
${\bepsilon}^{\rm in}$ and ${\bepsilon}^{\rm out}$,
respectively. 
Then the effective photon-phonon interaction
for the phonon mode ${\bf A}^{\rm q}$ 
is given by
\begin{align}
 {\cal H}_{\rm G} 
 \equiv e^2({\bepsilon}^{\rm out} \cdot {\bv}\sigma_z )^\dagger
 ({\bf A}^{\rm q}\cdot {\bv}^{\rm q})
 ({\bepsilon}^{\rm in} \cdot  {\bv}\sigma_z).
 \label{eq:bulkint}
\end{align}
Here, we have omitted to write 
the electron propagator 
by assuming a resonance Raman process, 
in which 
the photo-excited electron is a resonant state.
Note that this effective interaction is for the bulk where
the electron propagator 
does not depend on the pseudospin.
Owing to Eq.~(\ref{eq:jqj}), ${\cal H}_{\rm G}$ 
can be expressed 
in terms of the axial velocity operator only.
Moreover, 
by using the properties of the axial velocity operator, 
$v_i^{\rm q}v_i^{\rm q} = \sigma_0 \tau_0$ ($i=x,y$)
and $v_x^{\rm q}v_y^{\rm q}=-v_y^{\rm q}v_x^{\rm q}$, 
we rewrite ${\cal H}_{\rm G}$ 
in the form~\footnote{A similar equation was derived
by Basko.~\cite{basko09} 
Note, however, that Eq.~(\ref{eq:hxy}) 
is different from Eq.~(53) in Ref.~\onlinecite{basko09} as 
the former contains the axial velocity $v_i^{\rm q}$, while 
the latter does not.
Due to $v_i^{\rm q}$, 
Eq.~(\ref{eq:hxy}) is symmetric with respect to the change 
of $x$ and $y$.
The existence of $v_i^{\rm q}$ in Eq.~(\ref{eq:hxy}) 
is essential in the subsequent arguments.}
\begin{align}
 \frac{{\cal H}_{\rm G}}{e^2} &= (\epsilon_x^{\rm in}\epsilon_x^{\rm out}-\epsilon_y^{\rm in}\epsilon_y^{\rm out})
 (-A_x^{\rm q}v_x^{\rm q}+A_y^{\rm q}v_y^{\rm q}) \nn \\
 & - (\epsilon_x^{\rm in}\epsilon_y^{\rm out}+\epsilon_y^{\rm in}\epsilon_x^{\rm out})
 (A_x^{\rm q}v_y^{\rm q}+A_y^{\rm q}v_x^{\rm q}).
 \label{eq:hxy}
\end{align}
Note that ${\cal H}_{\rm G}$
is linear in $v_i^{\rm q}$. 
This feature is unique to the G band
and is not seen in the case of the D and 2D bands
as we will show later.
By introducing the angles, $\theta_{\rm in}$, 
$\theta_{\rm out}$, and $\varphi_s$ as shown in
Fig.~\ref{fig:strain}, one has 
$\epsilon_x^{\rm in}=\epsilon^{\rm in}\cos(\theta_{\rm in}+\varphi_s)$,
$\epsilon_y^{\rm in}=\epsilon^{\rm in}\sin(\theta_{\rm in}+\varphi_s)$,
$\epsilon_x^{\rm out}=\epsilon^{\rm out}\cos(\theta_{\rm out}+\varphi_s)$,
$\epsilon_y^{\rm out}=\epsilon^{\rm out}\sin(\theta_{\rm out}+\varphi_s)$,
and 
\begin{align}
\begin{split}
 & A_x^{\rm q} = A_+^{\rm q} \cos \varphi_s - A_-^{\rm q} \sin\varphi_s, \\
 & A_y^{\rm q} = A_+^{\rm q} \sin \varphi_s + A_-^{\rm q} \cos\varphi_s. 
\end{split}
\label{eq:coordinate}
\end{align}
By putting these into Eq.~(\ref{eq:hxy}), we get 
\begin{align}
 \frac{{\cal H}_{\rm G}}{e^2 \epsilon^{\rm in}\epsilon^{\rm out}} = 
 &-(v_x^{\rm q} A_+^{\rm q} - v_y^{\rm q} A_-^{\rm q})
 \cos (\theta_{\rm in} + \theta_{\rm out}+\varphi_s) \nn \\
 &-(v_y^{\rm q} A_+^{\rm q} + v_x^{\rm q} A_-^{\rm q})
 \sin (\theta_{\rm in} + \theta_{\rm out}+\varphi_s).
\end{align}
The probability amplitude of the process is given by 
the expectation value
$\langle {\cal H}_{\rm G} \rangle$, and 
the Raman intensity of the $A_+^{\rm q}$ ($A_-^{\rm q}$) mode
is given by the square of the coefficient of 
$A_+^{\rm q}$ ($A_-^{\rm q}$) in 
$\langle {\cal H}_{\rm G} \rangle$ as 
\begin{align}
\begin{split}
 & I_{{\rm G}^+} \propto
 |\langle v_x^{\rm q}\rangle \cos(\Psi) + 
 \langle v_y^{\rm q}\rangle \sin(\Psi)|^2, \\
 & I_{{\rm G}^-} \propto
 |\langle v_y^{\rm q}\rangle \cos(\Psi) -
 \langle v_x^{\rm q}\rangle \sin(\Psi)|^2,
\end{split}
 \label{eq:IGpn}
\end{align}
where we have defined 
$\Psi \equiv \theta_{\rm in}+\theta_{\rm out}+\varphi_s$.
\footnote{In literature, $3\varphi_s$ 
instead of $\varphi_s$ appears in the angle $\Psi$. 
The factor of $3$ in front of $\varphi_s$
seems to come from the choice of
the coordinate system, 
for which $\varphi_s$ in Eq.~(\ref{eq:coordinate})
is replaced with $-\varphi_s$.~\cite{mohiuddin09} 
We could not find any proper reason for this choice.}

It is important to recognize that
only when $\langle v_y^{\rm q}\rangle=0$, 
$I_{{\rm G}^+}$ and $I_{{\rm G}^-}$
can have the crystallographic orientation dependence 
of the Raman intensity:
\begin{align}
 \begin{split}
 & I_{{\rm G}^+} \propto \cos^2(\Psi), \\
 & I_{{\rm G}^-} \propto \sin^2(\Psi),
\end{split}
\end{align}
which were used to fit the observed 
polarization dependence of the Raman intensity 
on the crystallographic orientation 
in strained graphene.~\cite{huang09,mohiuddin09} 
Note that without some constraint for ${\bv}^{\rm q}$,
the Raman intensity cannot 
have a crystallographic orientation dependence 
since the electronic dispersion is isotropic about the Dirac
point.~\footnote{This feature of graphene bulk is different from
that of nanotube because the electronic dispersion is not isotropic in
the case of nanotube due to the cutting lines.} 
The wave function in a periodic graphene 
does not yield a constraint for the axial velocity ${\bv}^{\rm q}$, 
so that not only $\langle v_x^{\rm q} \rangle$
but also $\langle v_y^{\rm q} \rangle$ 
can take a nonzero value.
In this case, we can have
$\langle v_x^{\rm q} \rangle= \cos\theta({\bf k})$
and $\langle v_y^{\rm q} \rangle=\sin\theta({\bf k})$, where
$\theta({\bf k})$ is the angle between the wave vector ${\bf k}$
and the $k_x$-axis.
After the integral over the variable $\theta({\bf k})$, 
the intensity becomes independent of the angle $\Psi$.
Since we have the constraint $\langle v_y^{\rm q} \rangle = 0$
for both the zigzag and armchair orientations,
it is naturally expected 
for the graphene sample with rectangle shape enclosed by zigzag and
armchair edges shown in Fig.~\ref{fig:unit}(b)
that we still have $\langle v_y^{\rm q} \rangle =0$, and 
that only $\langle v_x^{\rm q} \rangle$ 
can have a non-vanishing value.
Then we can reproduce the crystallographic orientation dependence.
As a matter of course, 
there remains a question of 
whether or not the constraint $\langle v_y^{\rm q} \rangle = 0$ holds
in a graphene sample with a general edge shape.
A further discussion on this point will be given in Sec.~\ref{sec:ds}.

If the Raman intensity of the G band 
in the bulk of graphene
does not have a polarization dependence,
there are in principle two ways to interpret this.
One way is to assume that
the bulk of graphene with edge is identical to the ``bulk'' of a periodic
graphene without edge.
In this case both $I_{{\rm G}^+}$ and $I_{{\rm G}^-}$
do not have any polarization dependence.
The other way is to assume that the existence of graphene edge 
gives rise to some constraint for the axial velocity 
in the ``bulk'' like $\langle v_y^{\rm q}\rangle=0$ with respect to 
the states participating in the Raman process.
In this case both $I_{{\rm G}^+}$ and $I_{{\rm G}^-}$
do have polarization dependence, but the sum of them 
$I_{{\rm G}^+}+I_{{\rm G}^-}$ does not.
Thus, without strain, 
the two kinds of ``bulk'' can not be distinct.
The experimental results~\cite{huang09,mohiuddin09} 
in the bulk of strained graphene indicate
that the later interpretation is plausible.

\subsection{Edge}

We now examine 
the Raman intensity of the ${\rm G}^+$ mode 
and that of ${\rm G}^-$ mode for the graphene edge.
Since we have the constraint 
$\langle v_y^{\rm q} \rangle = 0$
for the standing wave, 
we need to modify Eq.~(\ref{eq:bulkint}) at the graphene edge 
as 
\begin{align}
 \langle {\cal H}'_{\rm G} \rangle 
 \equiv e^2 \langle ({\bepsilon}^{\rm out} \cdot {\bv}\sigma_z )^\dagger \rangle
 \langle {\bf A}^{\rm q}\cdot {\bv}^{\rm q} \rangle
 \langle {\bepsilon}^{\rm in} \cdot  {\bv}\sigma_z \rangle.
 \label{eq:edgeint}
\end{align}
In contrast to the effective interaction for the bulk
given in Eq.~(\ref{eq:bulkint}),
each interaction operator is replaced with the expectation value of
the operator, by which the intermediate state can satisfy 
the constraint for the standing wave.
Physically speaking, this modification 
assumes that the coherence between 
ingoing and outgoing states of the standing wave
is strong, so that the intermediate state can not transfer into 
an ingoing state or an outgoing state independently.
This coherence may be weak in the bulk, for which case 
Eq.~(\ref{eq:bulkint}) would become a better approximation.
We note that regardless of the weakness of the coherence in the bulk,
the coherence for the initial and final states
in the Raman process can lead to the crystallographic orientation
dependences of the ${\rm G}^+$ and ${\rm G}^-$ bands.

Now, with the constraint $\langle v_y^{\rm q}\rangle = 0$, 
Eq.~(\ref{eq:edgeint}) becomes 
\begin{align}
 \langle {\cal H}'_{\rm G} \rangle = 
 e^2 A_x^{\rm q} \epsilon^{\rm out}_y \epsilon^{\rm in}_y
 \langle v_x^{\rm q} \rangle^3.
 \label{eq:Gpoledge}
\end{align}
This is a mathematical expression of the selection rule for
the G band near the graphene edge in shortened form.
Since 
$\epsilon_y^{\rm in}=\epsilon^{\rm in}\sin(\theta_{\rm in}+\varphi_s)$, and
$\epsilon_y^{\rm out}=\epsilon^{\rm out}\sin(\theta_{\rm
out}+\varphi_s)$
(see Fig.~\ref{fig:strain}),
we have with Eq.~(\ref{eq:coordinate}) that
$\langle {\cal H}'_{\rm G} \rangle\propto
 (A_+^{\rm q}\cos\varphi_s - A_-^{\rm q}\sin\varphi_s)
 \sin(\theta_{\rm in}+ \varphi_s)
 \sin(\theta_{\rm out}+ \varphi_s)$.
From the coefficients of $A_+^{\rm q}$ and $A_-^{\rm q}$
in this representation,
the Raman intensity of each mode is given by 
\begin{align}
 \begin{split}
  & I_{{\rm G}^+} \propto \cos^2(\varphi_s)
  \sin^2(\theta_{\rm in}+ \varphi_s)
 \sin^2(\theta_{\rm out}+ \varphi_s), \\
  & I_{{\rm G}^-} \propto \sin^2(\varphi_s)
  \sin^2(\theta_{\rm in}+ \varphi_s)
 \sin^2(\theta_{\rm out}+ \varphi_s).
 \end{split}
 \label{eq:sGint}
\end{align}
It is amusing to note that 
$I_{{\rm G}^-}$ ($I_{{\rm G}^+}$) vanishes
when $\varphi_s =0$ ($\varphi_s =90^\circ$).
Note also that the ratio $I_{{\rm G}^-}/I_{{\rm G}^+}$
depends only on the angle $\varphi_s$, which 
is in contrast to the case of the bulk. 

\subsection{Kohn Anomaly Effect}

\begin{figure}[htbp]
 \begin{center}
  \includegraphics[scale=0.5]{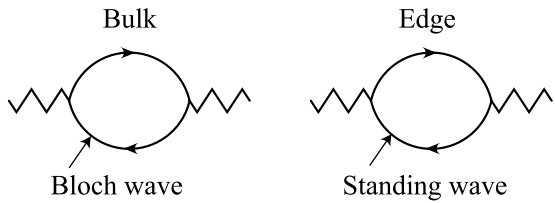}
 \end{center}
 \caption{Kohn anomaly effect at the bulk and edge.
 In the bulk, the intermediate electron-hole state 
 consists of the Bloch plane wave, while at the edge, 
 the intermediate state consists of the standing wave.
 The ``spin'' of the Bloch state can point arbitrary direction, while
 that of the standing wave can not have a component parallel to 
 $\sigma_y \tau_z$.
 }
 \label{fig:loop}
\end{figure}

Kohn anomaly effect is useful for illuminating
the essential difference between 
the predictions of the two models 
for the bulk and edge [Eqs.~(\ref{eq:bulkint}) and (\ref{eq:edgeint})].
For the G band, 
the Kohn anomaly effect is caused by 
the electron-hole pair creation 
from the phonon 
which is described as a vertical transition 
in the picture of the Dirac cone.~\cite{lazzeri06prl,ando06-ka,yan07}
The probability amplitude for the vertical pair creation 
from a phonon mode ${\bf A}^{\rm q}$ is given by 
\begin{align}
 M_{\rm G} = \langle ({\bf A}^{\rm q} \cdot \bv^{\rm q}\sigma_z)^\dagger
 ({\bf A}^{\rm q} \cdot \bv^{\rm q}\sigma_z) \rangle
 \label{eq:mg}
\end{align}
This is the formula for the bulk where
we assume that the ``spin'' 
(pseudospin and valleyspin)
of the intermediate state can be arbitrary 
(See Fig.~\ref{fig:loop}).
In other words, the propagator of electron
in the bulk is proportional to 
the identity matrix.
In this case, 
regardless of the character of the initial state,
we obtain $M_{\rm G} = {\bf A}^{\rm q}\cdot {\bf A}^{\rm q}$
from Eq.~(\ref{eq:mg}).
This means that both the LO and TO modes
in the bulk undergo the Kohn anomaly effect. 
On the other hand, the corresponding matrix element 
for the edge is given by 
\begin{align}
 M'_{\rm G} = 
 \langle ({\bf A}^{\rm q} \cdot \bv^{\rm q}\sigma_z)^\dagger \rangle 
 \langle {\bf A}^{\rm q} \cdot \bv^{\rm q}\sigma_z \rangle.
\end{align}
For the case of edge, 
we assume that the intermediate state
is given by the standing wave which has 
the constraint condition for the ``spin''.
It is easy to show that
$M'_{\rm G} = ({\bf A}^{\rm q}\times \langle \bv \rangle)^2$. 
This leads to the selection rule for the G band at edge
since we have $\langle v_y^{\rm q} \rangle = 0$.
This is the cause of the asymmetry that 
only the LO mode can undergo
the Kohn anomaly effect at both the zigzag and armchair
edges.~\cite{sasaki09,sasaki10-jpsj}

\section{D and 2D Bands}\label{sec:Dand2D}

In this section 
we study the polarization dependences
of the Raman intensities of the 
D and 2D (G$'$) bands for the bulk and edge.
It is shown that 
the Raman intensity of the D band in the bulk 
does not have a polarization dependence, 
while that of the 2D band in the bulk 
can have the polarization dependence, 
$I_{\rm 2D} \propto (\bepsilon^{\rm in}\cdot\bepsilon^{\rm out})^2$.
It is also shown that the Raman intensities of the D and 2D bands 
in strained graphene do not have a crystallographic orientation
dependence in the bulk. 
At the edge, these bands can exhibit 
the polarization dependence 
similar to that of the G band and also 
have the crystallographic orientation dependence.

The off-diagonal term in Eq.~(\ref{eq:Hami}),
$\phi \sigma_x$ (or $\phi^*\sigma_x$), 
represents intervalley phonon modes 
which are responsible for the Raman D and 2D bands.
Although the D and 2D bands consist of several phonon modes with
different wave vectors
which depend on the excitation laser energy, 
we examine the Kekul\'e distortion
as the representative mode. 
It is straightforward 
to show that 
$\phi$ is a constant for Kekul\'e distortion,
and $|\phi|$ is about three times larger than 
$|{\bf A}^{\rm q}|$ for the G band.~\cite{sasaki08ptps}
The latter can explain why the intensity 
of the 2D band is much larger than that
of the G band.
Let us denote $\phi = e^{i\theta} |\phi|$, 
then the el-ph interaction for the D band 
$H_{\rm D}$ is written by 
\begin{align}
 H_{\rm D} = |\phi| \sigma_x \tau_\theta,
 \label{eq:HD}
\end{align}
where $\tau_\theta \equiv \tau_x \cos\theta - \tau_y \sin\theta$.
Because the matrices $\sigma_x \tau_x$ and $\sigma_x \tau_y$ 
do not appear in ${\bv}$ and ${\bv}^{\rm q}$,
the D band can give us 
new information on the electronic structure
that is not included in the G band.
We will leave the phase $\theta$ of $\phi$ unspecified
because $\theta$ relates to the TO modes
near the $\Gamma$ point through a gauge transformation
as is shown in Sec.~\ref{ssec:gt}.

\subsection{D Band}

The effective photon-phonon interaction 
for the Kekul\'e mode is written by
\begin{align}
 {\cal H}_{\rm D} 
 \equiv |\phi| ({\bepsilon}^{\rm out} \cdot {\bv}\sigma_z )^\dagger
 (\sigma_x \tau_\theta)
 ({\bepsilon}^{\rm in} \cdot  {\bv}\sigma_z).
\end{align}
It is easy to find that 
\begin{align}
 \langle {\cal H}_{\rm D} \rangle 
 \propto {\bepsilon}^{\rm in}\cdot {\bepsilon}^{\rm out}
 \langle \sigma_x \tau_\theta \rangle
 - i {\bepsilon}^{\rm in} \times {\bepsilon}^{\rm out}
 \langle \sigma_y \tau_{\theta+\frac{\pi}{2}} \rangle.
\end{align}
Hence, when ${\bepsilon}^{\rm in}$ 
is parallel with ${\bepsilon}^{\rm out}$, i.e., 
when ${\bepsilon}^{\rm in} \times {\bepsilon}^{\rm out}=0$,
the D band intensity is proportional to 
$|\langle \sigma_x \tau_\theta \rangle|^2$. 
On the other hand, 
when ${\bepsilon}^{\rm in}$ 
is perpendicular to ${\bepsilon}^{\rm out}$, i.e., 
${\bepsilon}^{\rm in} \cdot {\bepsilon}^{\rm out}=0$,
the D band intensity is proportional to 
$|\langle \sigma_y \tau_{\theta+\frac{\pi}{2}} \rangle|^2$.
If $|\langle \sigma_x \tau_\theta \rangle|\ne 
|\langle \sigma_y \tau_{\theta+\frac{\pi}{2}} \rangle|$,
the D band intensity in the bulk
can have a polarization dependence.
However, we could not find 
any special reason for this asymmetry.
Rather, it is probable that 
$|\langle \sigma_x \tau_\theta \rangle|=
|\langle \sigma_y \tau_{\theta+\frac{\pi}{2}} \rangle|$
holds in the bulk.
It is reasonable to consider that 
the D band intensity does not have 
a polarization dependence in the bulk.

Since the zigzag edge is not 
the source of intervalley scattering,
we have $\langle \sigma_i \tau_x \rangle = 0$ 
and $\langle \sigma_i \tau_y \rangle = 0$
for the standing wave near the zigzag edge.
Thus, we get 
\begin{align}
 \langle H_{\rm D} \rangle= 0.
\end{align}
This shows that 
the D band intensity is suppressed 
near the zigzag edge.
In contrast, the armchair edge 
is the source of intervalley scattering.
In fact, the standing wave near the armchair edge is given by 
\begin{align}
 \Psi_{\bf k}({\bf r}) = C e^{ik_y y}\Phi_{\bf k}
\begin{pmatrix}
  e^{+ik_x x} \cr e^{ia} e^{-ik_x x}
\end{pmatrix},
\label{eq:wfarm}
\end{align}
where $C$ is normalization constant and $\Phi_{\bf k}$ is the wave
function of the pseudospin.~\cite{sasaki10-jpsj,sasaki10-chiral}
It is easy to show that 
this wave function reproduces 
$\langle v_x \rangle =0$ and 
$\langle v_y^{\rm q} \rangle =0$, 
which is consistent with the results obtained in Sec.~\ref{sec:selection}. 
Moreover, 
pseudospin and valleyspin can be calculated separately
since the armchair edge preserves the pseudospin.
From Eq.~(\ref{eq:wfarm}), we get 
\begin{align}
\begin{split}
 & \langle \sigma_x \tau_x \rangle = 
 \langle \sigma_x \rangle \cos(a-2k_x x), \\ 
 & \langle \sigma_x \tau_y \rangle = 
 \langle \sigma_x \rangle \sin(a-2k_x x). 
\end{split} 
\end{align}
By using these results, 
we obtain 
\begin{align}
 \langle H_{\rm D} \rangle = |C|^2 \int_S |\phi| \langle \sigma_x \rangle 
 \cos \left( \theta + a - 2k_x x \right)dxdy,
\end{align}
for the armchair edge.
To summarize, the D band has an obvious selection rule;
the D band is Raman active at the armchair edge 
while it is not active at the zigzag edge. 
This has been a well-known fact which is useful 
in distinguishing armchair-dominated edge 
from zigzag-dominated edge.~\cite{malard09,canifmmode04sec,you08,gupta09}

The polarization dependence of the D band intensity
near the armchair edge
is different from that in the bulk. 
In fact, we get from 
$\langle {\cal H}'_{\rm D} \rangle
\equiv |\phi| \langle {\bepsilon}^{\rm out} \cdot {\bv}\sigma_z \rangle^\dagger
\langle \sigma_x \tau_\theta \rangle
\langle {\bepsilon}^{\rm in} \cdot  {\bv}\sigma_z \rangle$
that
\begin{align}
 \langle {\cal H}'_{\rm D} \rangle
 \propto |\phi|\epsilon_y^{\rm in}\epsilon_y^{\rm out} 
 \langle v_x^{\rm q} \rangle^3.
 \label{eq:Dpoledge}
\end{align}
Let the angle 
between the armchair edge and the polarization
of the incident (scattered) laser be 
$\Theta_{\rm in}$ ($\Theta_{\rm out}$).
Then we can use
$\epsilon_y^{\rm in} = \epsilon^{\rm in}\cos\Theta_{\rm in}$
and 
$\epsilon_y^{\rm out} = \epsilon^{\rm out}\cos\Theta_{\rm out}$
in Eq.~(\ref{eq:Dpoledge}).
Thus, the D band intensity at the armchair edge 
behaves according to 
$I_{\rm D} \propto \cos^2\Theta_{\rm in}\cos^2\Theta_{\rm out}$
which is maximum when the polarization of the incident (or scattered)
light is parallel to the edge.


Here, let us mention 
experiments on the polarization dependence
for the D band intensity at the edge.
First, the polarization dependence of 
$I_{\rm D} \propto \cos^2\Theta_{\rm in}\cos^2\Theta_{\rm out}$
is consistent with the observation 
for graphite edges by
Can\ifmmode \mbox{\c{c}}\else \c{c}\fi{}ado {\it et
al.}~\cite{canifmmode04sec,canifmmode04} and
the observations 
for edges of single-layer graphene by 
You {\it et al.}~\cite{you08}, 
Gupta {\it et al.}~\cite{gupta09} and 
Casiraghi {\it et al.}~\cite{casiraghi09}
Secondly, for edges of single-layer graphene,
Cong {\it et al.}~\cite{cong10} 
confirmed that not only the D band but also 
the G band follows $I_{\rm D,G} \propto \cos^2\Theta_{\rm in}$
for the polarization of the incident laser light, 
whereas You {\it et al.}~\cite{you08}
and Gupta {\it et al.}~\cite{gupta09} did not observe 
any polarization dependence for the G band.
On the other hand, 
for edges of bilayer graphene,
Begliarbekov {\it et al.}~\cite{begliarbekov10}
showed that the G band intensity 
had the polarization dependence, 
while the D band intensity 
did not exhibit any polarization dependence. 
The polarization dependence of the D band 
at the armchair edge might be sensitive to 
the number of graphene layers.
In fact, Gupta {\it et al.}~\cite{gupta09} shows that 
three Lorentzian components are necessary to fit 
the Raman spectrum of the D band in bilayer graphene, whereas
the D band spectra 
in a single-layer graphene can be well fitted by 
a single Lorentzian component.
Theoretically, 
by comparing Eq.~(\ref{eq:Dpoledge}) with Eq.~(\ref{eq:Gpoledge}),
we see that the polarization dependence of the D band intensity 
at the armchair edge
is identical to that of the G band intensity 
at the armchair edge.
We also note that 
when zigzag and armchair edges are randomly 
distributed along a mixed edge, 
the G band does not 
show a polarization dependence.~\cite{sasaki10-jpsj} 
However, even in this case, 
the D band should have a polarization dependence
since there is no counterpart of the D band which 
can erase the polarization dependence.
In the case of the G band, two components (LO and TO modes)
can coexist in the random edge. 
They have different polarization dependence, so that 
a polarization dependence of the G band may diminish
in the case of a random edge.

Equation~(\ref{eq:Dpoledge}) 
leads to the following polarization 
dependence on the crystallographic orientation 
in strained graphene,
\begin{align}
 I_{{\rm D}} \propto 
 \sin^2(\theta_{\rm in}+ \varphi_s)
 \sin^2(\theta_{\rm out}+ \varphi_s).
\end{align}
Thus, from Eq.~(\ref{eq:sGint}) we find that 
$I_{\rm D} \propto I_{{\rm G}^+}+I_{{\rm G}^-}$.
This might be one of the most interesting consequence 
for the G and D bands concerning with the armchair edge
in strained graphene.

\subsection{2D Band}

The effective el-ph interaction for the 2D band is 
given by the square of $H_{\rm D}$ as
\begin{align}
 H_{\rm 2D}= 
 H_{\rm D}^2 
 = |\phi|^2 \sigma_0 \tau_0.
\end{align}
Because the effective interaction 
is proportional to the identity matrix $\sigma_0\tau_0$,
no constraint can affect the el-ph matrix element
for the 2D band.
The effective photon-phonon interaction for the 2D band
in the bulk is given by
\begin{align}
 {\cal H}_{\rm 2D} 
 \equiv ({\bepsilon}^{\rm out} \cdot {\bv}\sigma_z )^\dagger
 (|\phi|^2 \sigma_0 \tau_0)
 ({\bepsilon}^{\rm in} \cdot  {\bv}\sigma_z).
\end{align}
Then we have 
\begin{align}
 \langle {\cal H}_{\rm 2D} \rangle
 \propto ({\bepsilon}^{\rm in} \cdot {\bepsilon}^{\rm out})\sigma_0 \tau_0
 - i ({\bepsilon}^{\rm in} \times {\bepsilon}^{\rm out}) \sigma_z \tau_z.
\end{align}
Note that $\langle \sigma_0 \tau_0 \rangle = 1$ holds 
for any kind of wave function. 
Furthermore, 
the condition $\langle \sigma_z \tau_z \rangle=0$ should be satisfied
in the absence of a magnetic field.~\cite{sasaki08ptps}
Consequently, the 2D band intensity follows 
$I_{\rm 2D} \propto ({\bepsilon}^{\rm in} \cdot {\bepsilon}^{\rm out})^2$
in the absence of a magnetic field.
The Raman intensity of the 2D band in bulk is maximum when the incident
and scattered polarizations are parallel and minimum when they are
orthogonal, which is in good agreement with the experimental result by
Yoon {\it et al.}~\cite{yoon08}
The polarization dependence of the 2D band 
closely resembles that of Rayleigh scattering~\cite{casiraghi07}
because the effective Hamiltonian for Rayleigh scattering  
is given by 
\begin{align}
 {\cal H}_{\rm R} 
 \equiv e^2 
 ({\bepsilon}^{\rm out} \cdot {\bv}\sigma_z )^\dagger
 ({\bepsilon}^{\rm in} \cdot  {\bv}\sigma_z).
\end{align}
Note that ${\cal H}_{\rm R}$ is the same as ${\cal H}_{\rm 2D}$
except for the numerical factor (coupling constant).
In contrast, for the edge, 
the polarization dependence of the 2D band intensity 
is the same as that of the D band:
\begin{align}
 \langle {\cal H}'_{\rm 2D} \rangle
 \propto |\phi|^2 \epsilon_y^{\rm in}\epsilon_y^{\rm out} 
 \langle v_x^{\rm q} \rangle^2.
\end{align}
Note, however, that 
because the el-ph matrix element for the 2D band
is given by $\langle H_{\rm 2D} \rangle=|\phi|^2$
regardless of the orientation of the edge,
the 2D band intensity appears both 
at the zigzag and armchair edges.
The constraint works for the optical transition only, and therefore the
polarization dependence of the 2D band follows the same rule for 
the G band. 

It is known that 
Eq.~(\ref{eq:HD}) does not cover the deformations
representing a pentagon or heptagon.~\cite{gonzalez92,lammert00}
For these topological defects, 
some combination of $\sigma_a \tau_b$
besides $\sigma_x \tau_x$ and $\sigma_x \tau_y$
can appear.
Thus, besides the appearance of the Raman peak 
due to the vibrational dynamics 
specific to the topological defect,~\cite{doyle95}
the D and 2D bands 
can have some information on the presence 
of the topological defect.
In fact, 
it is known that a single pentagon or a single heptagon 
gives rise to a mixing between K and K$'$ points, leading to a
sophisticated topological effect on the wave function.
Recently, the existence of 
a new type of graphene edge called reczag 
(reconstructed zigzag) has been proposed.~\cite{koskinen09}
Note that 
this reczag edge consists of a pair of pentagon and heptagon
along the edge.
In this case, the topological effect 
on the wave function is not significant
because the topological effect of a single pentagon
is cancelled by that of a single heptagon.
In fact, a numerical calculation shows 
the appearance of the edge states
near the reczag edge.~\cite{koskinen08}
This indicates that the standing wave near the reczag edge
is similar to that near the usual zigzag edge.

\subsection{Gauge Transformation and D$'$ Band}\label{ssec:gt}

At first sight, due to the momentum conservation,
a TO mode with small nonzero momentum 
(${\bf q}\ne 0$ and $|{\bf q}| \ll |{\bf k}_{\rm F}|$)
may cause an intravalley scattering, 
but is not expected to be relevant to
an intervalley scattering.
Here, in terms of the gauge transformation,
we shall show that such TO modes
do not contribute to intravalley scattering; rather 
they can be activated through the intervalley scattering.

The TO modes with small momentum 
can be represented by the derivative of some scalar function 
$\varphi({\bf r})$ as 
\begin{align}
 {\bf A}^{\rm q}_{\rm TO}({\bf r}) = 
 {\bf A}^{\rm q}_{\rm TO} + \nabla \varphi({\bf r}),
\end{align}
where ${\bf A}^{\rm q}_{\rm TO}$ on the right-hand side 
is the zero mode which has been relevant to the G
band.~\cite{sasaki08_curvat} 
Due to the following gauge transformation, 
the scalar function can be transferred 
to the phases of the wave function and $\phi$
as 
\begin{align}
 \begin{pmatrix}
  \bsigma \cdot (\hat{{\bf p}}+{\bf A}^{\rm q}_{\rm TO}) & 
  \tilde{\phi} \sigma_x \cr
  \tilde{\phi}^* \sigma_x & 
  \bsigma' \cdot (\hat{{\bf p}}-{\bf A}^{\rm q}_{\rm TO})
 \end{pmatrix}
 \begin{pmatrix}
  \tilde{\Psi}_{\rm K} \cr \tilde{\Psi}_{\rm K'}
 \end{pmatrix},
\end{align}
where $\tilde{\Psi}_{\rm K} = e^{-i\varphi({\bf r})} \Psi_{\rm K}$,
$\tilde{\Psi}_{\rm K'} = e^{+i\varphi({\bf r})} \Psi_{\rm K'}$, and
$\tilde{\phi} = e^{-2i\varphi({\bf r})} \phi$.
Note that the TO mode appears as the phase of $\phi$ 
(see $\theta$ in Eq.~(\ref{eq:HD})).
A physical significant of this gauge transformation 
is that the TO mode can be excited in combination with 
the intervalley scattering.
Since the armchair edge enhances the intervalley scattering
$\phi$, the TO mode with small nonzero momentum 
$\varphi({\bf r})$ 
also can be excited near the armchair edge.

The LO modes with small momentum 
can be represented by the derivative of 
some scalar function $\chi({\bf r})$ as 
\begin{align}
 {\bf A}^{\rm q}_{\rm LO}({\bf r}) = 
 {\bf A}^{\rm q}_{\rm LO} + \nabla \times (\chi({\bf r}){\bf e}_z).
\end{align}
In contrast to the TO mode, 
${\bf A}^{\rm q}_{\rm LO}({\bf r})$ can not be gauge transformed 
into the phase of $\phi$ 
because it has a non-vanishing field strength: 
$B_z^{\rm q}= \nabla \times {\bf A}^{\rm q}_{\rm LO}({\bf r})\ne 0$.~\cite{sasaki05}
Thus, these LO modes are responsible for intravalley scattering.
The D$'$ band~\cite{saito01prl} 
observed slightly above the G band in Raman spectra
(around 1620${\rm cm}^{-1}$)
is originated from these LO modes.
The Hamiltonian for the D$'$ band is given by
\begin{align}
 H_{\rm D'} =
 {\bf A}^{\rm q}({\bf r})\cdot \bv^{\rm q}.
\end{align}
Note that the ``spin'' structure is 
the same as that for the G band.
Thus, the polarization dependence of the D$'$ band
follows that of the G band in the bulk. 
At edge, we have
$\langle H_{\rm D'} \rangle=
 \int_S A_x^{\rm q}({\bf r})\Psi^\dagger({\bf r}) v^{\rm q}_x \Psi({\bf r})dxdy$.
The intensity behaves as 
$I_{\rm D'} 
\propto |\epsilon^{\rm out}_y \epsilon^{\rm in}_y|^2$.
The polarization dependence of the D$'$ band is the same as that of the
G band at edge, however, it might be difficult to observe 
the polarization dependence of $I_{\rm D'}$
due to its small intensity.

\begin{table*}[htbp]
 \caption{\label{tab:1} 
 Polarization dependences of the Raman intensities for the
 optical phonon modes in strained graphene. 
 $\times$ represents absence of a polarization dependence. 
 For ``Bulk'', $\theta_{\rm in}$ ($\theta_{\rm out}$) denotes 
 the angle between the strain and the incident (scattered)
 laser polarization, and 
 $\varphi_s$ is the angle between the direction of strain and the zigzag
 edge ($x$-axis).
 For the ``Armchair'' and ``Zigzag'' edges, 
 $\Theta_{\rm in}$ ($\Theta_{\rm out}$) denotes 
 the angle between the incident (scattered) laser polarization
 and the armchair or zigzag edge. 
 The polarization dependence of the G band Raman intensity 
 in unstrained graphene is given by $I_{G^+}+I_{G^-}$, so that 
 the polarization dependence on the crystallographic orientation of
 strained graphene is lost at the bulk. 
 }
 \begin{ruledtabular}
  \begin{tabular}{c|cccc}
   & G$^+$ & G$^-$ & D & 2D (G$'$)  \\
   \hline 
   & \\
   Bulk & $\cos^2 (\theta_{\rm in}+\theta_{\rm out}+\varphi_s)$ &
	   $\sin^2 (\theta_{\rm in}+\theta_{\rm out}+\varphi_s)$ &
	       $\times$ & $(\bepsilon^{\rm in}\cdot \bepsilon^{\rm
		   out})^2 $  \\
   & \\
   \hline   
   & \\
   Armchair & $\cos^2(\varphi_s)\cos^2(\Theta_{\rm in})\cos^2(\Theta_{\rm out})$ & 
	   $\sin^2(\varphi_s)\cos^2(\Theta_{\rm in})\cos^2(\Theta_{\rm out})$ & 
	       $\cos^2(\Theta_{\rm in})\cos^2(\Theta_{\rm out})$ &
		   $\cos^2(\Theta_{\rm in})\cos^2(\Theta_{\rm out})$ \\
   & \\
   \hline 
   & \\
   Zigzag & $\cos^2(\varphi_s)\sin^2(\Theta_{\rm in})\sin^2(\Theta_{\rm
       out})$ & $\sin^2(\varphi_s)\sin^2(\Theta_{\rm
	   in})\sin^2(\Theta_{\rm out})$ & $\times$ &
		   $\sin^2(\Theta_{\rm in})\sin^2(\Theta_{\rm out})$ \\
   & \\
  \end{tabular}
 \end{ruledtabular}
\end{table*}

\section{Discussion and Summary}\label{sec:ds}

Here, we would like to mention 
the status of the experiment 
on the G band for the graphene edge.
Cong {\it et al.}~\cite{cong10} 
conducted a systematic research 
on edges of single layer graphene
and found, in particular, that 
there were two orientations of the graphene edge 
(A-edge and Z-edge) which 
exhibited different behaviors 
against the polarization of the incident laser light.
The Raman intensity of the A-edge ($I_{\rm A}$) is enhanced 
when the polarization becomes parallel to the edge
and that of the Z-edge ($I_{\rm Z}$) is enhanced 
when the polarization becomes perpendicular to the edge.
By using the angle $\Theta$ 
between the orientation of the edge
and the polarization of the incident laser light, 
they could fit 
the observed Raman intensities 
$I_{\rm A}(\Theta)$ and $I_{\rm Z}(\Theta)$ 
with
\begin{align}
 \begin{split}
  & I_{\rm A}(\Theta) = a + b \cos^2 \Theta, \\
  & I_{\rm Z}(\Theta) = c + d \sin^2 \Theta,
 \end{split}
\end{align}
where $a$, $b$, $c$, and $d$ are fitting parameters.
In their experimental data, 
the maximum intensity ($a+b$ or $c+d$)
is about two times larger than the minimum intensity
($a$ or $c$), 
so that $a/(a+b)$ and $c/(c+d)$ is about $1/2$.

The appearance of these two behaviors for the G band 
is consistent with the selection rule at the graphene
edge.~\cite{sasaki10-jpsj}
The A-edge is considered to be armchair dominant edge and
the Z-edge is zigzag dominant edge.
The ratio of the minimum intensity 
to the maximum intensity corresponds to the 
square of the ratio of the zigzag (armchair) part 
to the armchair (zigzag) part 
in a mixed edge.~\cite{sasaki10-jpsj}
We thus estimate that the A-edge consists of 
60$\%$ armchair and 40$\%$ zigzag edge, while the Z-edge
consists of 40$\%$ armchair and 60$\%$ zigzag edge.
We can get the similar value for the data obtained by 
Begliarbekov {\it et al.}~\cite{begliarbekov10}
who carried out polarization resolved micro-Raman spectroscopy 
at edges of bilayer graphene.
The two orientations (A-edge and Z-edge) were clearly resolved
even in bilayer graphene, which also suggests that 
the number of graphene layers 
does not invalidate the selection rule
for the G band.
Can\ifmmode \mbox{\c{c}}\else \c{c}\fi{}ado {\it et
al}.~\cite{canifmmode04}
observed that the Raman intensity of the G band for a nanoribbon
located on top of a highly oriented pyrolytic graphite (HOPG)
has a strong dependence on the incident light
polarization.
They showed that the Raman intensity is maximum when 
the polarization is parallel to the edge of a nanoribbon. 
Their result is consistent with 
the selection rule for the armchair edge.
A notable point in their experiment is that 
the ratio of the maximum intensity to the minimum intensity 
was very high. 
We speculate that the nanoribbon located on top of HOPG
had rather regular armchair edge.

At this moment, we do not know 
how to make a clear distinction between bulk and edge. 
In other words, 
there exists no criteria 
by which we can decide 
whether the Raman process is best described by
$\langle {\cal H}_{\rm G} \rangle$ (Eq.~(\ref{eq:bulkint}))
or $\langle {\cal H}'_{\rm G} \rangle$ (Eq.~(\ref{eq:edgeint})).
Since the electron dynamics in graphene 
is given by massless Dirac equation which is a scale-less theory,
we consider that 
bulk of graphene can not be completely separated from 
the edge.
To put it in an extreme way,
there is no bulk region in a nanoribbon with perfect regular edge, 
as well as that there is no edge region in a nanotube. 
The problem is the case of a mixed rough edge which 
might bring 
a characteristic length scale to the scale-less theory.
Theoretical estimation of the effective length 
is an important issue should be carried out
in the near future. 
Experimentally, 
according to the Raman mapping data by Cong {\it et al.},~\cite{cong10} 
the effective region from the edge in which 
the description using the standing wave is valid,
is about 400 nm, which seems to be comparable to 
the Gaussian laser beam waist.~\cite{gupta09}

Let us investigate the property of an eigenstate 
in the interior part of graphene.
Since a graphene has the edge, 
the electronic wave function 
is given by the standing wave.
Furthermore, the expectation value of the velocity, 
$\langle {\bv} \rangle$,
must vanish, so that we have
$\langle v_x \rangle = 0$ and $\langle v_y \rangle = 0$
for the standing wave.
Note that in the case of nanotubes
the velocity around the axis of the tube 
takes nonzero value in general.
Suppose that the graphene is surrounded by the zigzag edges only
[See Fig.~\ref{fig:strain}(b,top)].
Then, 
$\langle v_x \rangle = 0$ and $\langle v_y \rangle = 0$
must be satisfied at each valley since the zigzag edge
is not the source of intervalley scattering.
Because the velocity and the axial velocity are related 
with each other by $\bv^{\rm q} = \bv \tau_z$,
we have $\langle \bv^{\rm q} \rangle = 0$.
It is amusing to note that
in this special case Eq.~(\ref{eq:IGpn}) 
suggests that the resonant G band intensity vanishes. 
In contrast, 
when graphene is surrounded by the armchair edges only
[See Fig.~\ref{fig:strain}(b,bottom)], 
we have $\langle v_x^{\rm q} \rangle \ne0$ and 
$\langle v_y^{\rm q} \rangle =0$
since the armchair edge 
is relevant (irrelevant) to the valleyspin (pseudospin).
Note that in both hexagonal graphenes, 
we have at least the condition
$\langle v_y^{\rm q} \rangle =0$.
It is our speculation based on the observation of 
several edge shapes that 
the constraint 
$\langle v_y^{\rm q} \rangle =0$ holds 
for graphene with a wider variety of edge shapes.

The formalism using the gauge fields for photon ${\bf A}$
and phonon ${\bf A}^{\rm q}$ might provide 
a new and fresh insight into Raman scattering, 
which otherwise well-studied subject.
In ordinary Raman spectroscopy, 
we irradiate a laser light ${\bf A}^{\rm in}$
onto a graphene sample and observe 
the inelastically scattered light ${\bf A}^{\rm out}$. 
This Raman process in unstrained graphene 
may be represented as 
\begin{align}
 {\bf A}^{\rm in} \rightarrow {\bf A}^{\rm q} + {\bf A}^{\rm out}.
 \label{eq:raman}
\end{align}
The left-hand side of ``$\rightarrow$'' shows the input 
and the right-hand side of it denotes the output.
Note that a phonon ${\bf A}^{\rm q}$
on the right-hand side 
is a kind of lattice deformation or 
an internal strain. 
Thus, in Raman spectroscopy,
by inputting a photon (electronic) signal,
one gets a signal of strain from graphene.
Let us consider a process represented by 
\begin{align}
 {\bf A}^{\rm q}_{\rm in} \rightarrow {\bf A} + 
 {\bf A}^{\rm q}_{\rm out},
 \label{eq:strain}
\end{align}
where ${\bf A}^{\rm q}$ 
represents an external strain. 
This process of Eq.~(\ref{eq:strain})
is given by replacing 
${\bf A}$ (${\bf A}^{\rm q}$)
with ${\bf A}^{\rm q}$ 
(${\bf A}$) in Eq.~(\ref{eq:raman}).
In this process, 
by inputting strain, one gets
an electronic output ${\bf A}$ from graphene, which
seems to be a prototypical process of strain engineering.
Now, the Raman process in strained
graphene~\cite{ni08,yu08,huang09,mohiuddin09,tsoukleri09,robinson09,frank10}
is expressed by 
\begin{align}
 {\bf A}^{\rm q}_{\rm bulk} + {\bf A}^{\rm in} \rightarrow 
 {\bf A}^{\rm q}_{\rm bulk} + {\bf A}^{\rm q} + {\bf A}^{\rm out},
 \label{eq:sgra}
\end{align}
where the strain is described by ${\bf A}^{\rm q}_{\rm bulk}$.
Considering that 
this is a process which may be recognized as the sum of 
Eqs.~(\ref{eq:raman}) and (\ref{eq:strain}),
it can be said that Raman spectroscopy 
in strained graphene is a small step toward 
strain engineering in graphene.

It is reasonable to consider that Eq.~(\ref{eq:raman})
represents a Raman process in graphene without edge.
Strictly speaking, 
a Raman process in a real (unstrained) graphene 
should be represented not by Eq.~(\ref{eq:raman})
but by 
\begin{align}
 {\bf A}^{\rm q}_{\rm edge} + {\bf A}^{\rm in} \rightarrow 
 {\bf A}^{\rm q}_{\rm edge} + {\bf A}^{\rm q} + {\bf A}^{\rm out},
 \label{eq:egra}
\end{align}
because the presence of the edge is represented by 
a local strain field ${\bf A}^{\rm q}_{\rm
edge}$:
the zigzag edge is represented 
by a local ${\bf A}^{\rm q}$ field 
parallel to the edge, 
while the armchair edge corresponds to 
a local ${\bf A}^{\rm q}$ field 
normal to the edge.~\cite{sasaki06jpsj,sasaki10-chiral} 
It is interesting to note that 
the direction of the ${\bf A}^{\rm q}_{\rm edge}$
is coincident with that of the Raman active phonon mode
${\bf A}^{\rm q}$ near the graphene edge. 
Considering that 
${\bf A}^{\rm q}_{\rm bulk}$ in Eq.~(\ref{eq:sgra}) 
(${\bf A}^{\rm q}_{\rm edge}$ in Eq.~(\ref{eq:egra}))
represents a global (local) strain, 
the Raman spectroscopy near the graphene
edge~\cite{canifmmode04sec,you08,gupta09,cong10,begliarbekov10}
is complementary to the Raman spectroscopy 
in the bulk of strained graphene.

In conclusion,
the polarization dependences 
of the G, D, and 2D Raman bands
at bulk and edge 
have been investigated theoretically 
with paying attention to the pseudospin and valleyspin 
of the standing wave.
Our results are summarized in TABLE~\ref{tab:1}.
The constraint for the axial velocity 
provided by the graphene edge 
is essential to the selection rule for each Raman band.
The selection rules of the G and D bands 
for graphene edge are consistent 
with the recent experimental results.
The coherence provided by the graphene edge 
seems to persist even in the bulk, by which 
we explain the recent experiments for strained graphene 
showing the crystallographic orientation dependences 
of the Raman intensities of the ${\rm G}^+$ and ${\rm G}^-$ bands.
This also suggests that 
the ``bulk'' of graphene can not be completely
free from the surrounding graphene edge, and that 
it is necessary to distinguish 
the bulk of graphene that is surrounded by edge 
from the bulk of a periodic graphene without edge.

\section*{Acknowledgments}

K.S. would like to thank P. Kim, C. Cong, and T. You.
This work was motivated by the discussion with them.
He also wishes to thank S. Mathew for useful discussions.
This work is supported by 
a Grant-in-Aid for Specially Promoted Research
(No.~20001006) from the Ministry of Education, Culture, Sports, Science
and Technology (MEXT).

\bibliographystyle{apsrev}

\end{document}